\documentclass[journal]{IEEEtran}
\usepackage{amsmath}
\usepackage{stmaryrd}{
\usepackage{amsfonts}
\usepackage{mathrsfs}
\usepackage{amssymb,color,balance,cite}
\usepackage[dvips]{graphicx}

\begin{document}

\title{ Uncoordinated  Frequency Shifts based  Pilot Contamination  Attack
Detection   }
\author{Weile Zhang and Hai Lin$^\dag$ \\
MOE Key Lab for Intelligent Networks and Network Security,
Xi'an Jiaotong University, Xi'an, China \\
$^\dag$Department of Electrical and Information Systems, Osaka
Prefecture University, Osaka, Japan \\
wlzhang@mail.xjtu.edu.cn, lin@eis.osakafu-u.ac.jp
}

 \maketitle

 \pagestyle{empty}
 \thispagestyle{empty}


\begin{abstract}
Pilot contamination attack is an important kind of active eavesdropping activity conducted by a malicious user during channel training phase.
In this paper, motivated by the fact that frequency asynchronism could introduce divergence of the transmitted pilot signals between intended user and attacker, we propose a new uncoordinated frequency shift (UFS) scheme for detection of pilot contamination attack in multiple antenna system.
  An attack detection algorithm is further developed based on source enumeration method.
Both the asymptotic detection performance analysis and numerical results are provided to verify the proposed studies. The results demonstrate that the proposed UFS scheme can achieve comparable detection performance as the existing superimposed random sequence based scheme, without sacrifice of  legitimate channel estimation performance.
\end{abstract}

  \begin{keywords}
  Physical layer security, pilot contamination attack, uncoordinated frequency shift (UFS).
 \end{keywords}

\section{Introduction}

As an effective way to protect wireless transmissions from being eavesdropped,  physical layer security has been drawing substantial research interests. Recent results in physical layer security methods require full or partial knowledge about  channel state information of the legitimate system, which is quite vulnerable to smart malicious attacks.
A typical example in a TDD system is the so-called pilot contamination attack from an active eavesdropper~\cite{Zhou12}. As illustrated in Fig. 1, the eavesdropper (Eve) wants to overhear the communication from the legitimate transmitter (Alice) to the intended receiver (Bob). During the reverse uplink training phase, Bob sends training (pilot) signal to Alice, and the latter performs legitimate channel estimation based on channel reciprocity.
Unfortunately,  during the training phase, the active eavesdropper Eve can also send the same pilot signals, thereby biasing the channel estimation at Alice. This not only degrades the signal reception quality at Bob but also leads to a significant signal leakage to Eve
during the subsequent downlink data transmission.

The issue of above pilot contamination was first noted in~\cite{Zhou12} and a few works have been further reported to detect such smart attacks. For example, the random pilot sequence was employed in~\cite{Kapetanovic} to detect the presence of attack.
A few number of detection schemes were studied in~\cite{Kapetanovic14} assuming different knowledge about the large-scale fading parameters.
A two-way training scheme for discriminatory channel estimation was proposed in~\cite{YangTC14} though a whitening-rotation based semiblind method.
Based on the Neyman-Pearson criterion, a number of detection methods were developed in~\cite{KangVTC15} under different assumptions about the channel and noise statistics.
 The authors in~\cite{Xiong} proposed an energy ratio detector, by exploring the asymmetry of received signal power levels at the transmitter Alice and legitimate receiver Bob in the presence of an attack.
 Another recent work in~\cite{TugnaitWCL} proposed superimposing a random sequence on the training sequence at the Bob, allowing use of source enumeration methods to detect attack. The idea was further extended to the multiuser TDD/SDMA uplink scenario~\cite{TugnaitICASSP16}.

All of the above detection works have assumed perfect frequency synchronization in the system.  However, carrier frequency offset (CFO) naturally exists due to frequency mismatch between  transceiver oscillators~\cite{Zhang13,Zhang15sub}. This means both the legitimate user Bob and the attacker Eve should first perform CFO estimation and oscillator frequency calibration, such that the carrier frequency of their training signals can be aligned to Alice.
Although the CFOs are expected to be completely eliminated for data transmission, however, they may be beneficial to the purpose of pilot contamination attack.
Note that even the same pilot signals are transmitted from Bob and Eve, the natural CFOs would bring in individual phase shifts and thus result in divergence of the transmitted signals.

\begin{figure}[t]
\begin{center}
\includegraphics[width=50mm]{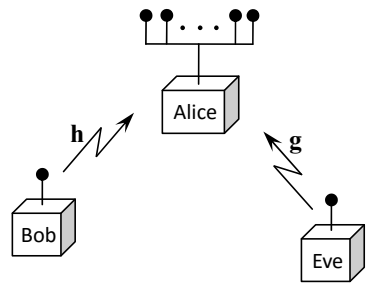}
\end{center}
\vspace{-3mm}
\caption{ Alice is a multi-antenna base station sending messages to the intended user Bob, while Eve is a malicious single-antenna eavesdropper. }
\end{figure}

Motivated by the above observation, we propose a new uncoordinated frequency shift (UFS) scheme for detection of pilot contamination attack in multiple antenna system. During the reverse training phase of the UFS scheme, Bob deliberately introduces multiple random frequency shifts when transmitting the publicly known pilot sequence. Eve has no knowledge about these random frequency shifts, and should be quite difficult to pretend exactly like Bob. This provides the opportunity to detect the presence of Eve. We further develop an attack detection algorithm based on source enumeration method.
Both the asymptotic detection performance analysis and the numerical results are provided to verify the proposed studies. The results demonstrate that the proposed UFS scheme can achieve comparable detection performance as the existing superimposed random sequence based scheme~\cite{TugnaitWCL}, but with a substantially improved legitimate channel estimation performance.

\textit{Notations:} Superscripts $(\cdot)^*$, $(\cdot)^T$ and $(\cdot)^H$ represent conjugate, transpose and Hermitian, respectively;
$E[\cdot]$ denotes expectation;
$\|\cdot\|$ denotes the Frobenius norm operator; $\mathbb{C}^{m\times n}$ defines the vector space of all $m\times n$ complex matrices;
${\bf j}=\sqrt{-1}$ is the imaginary unit;
$\textrm{diag}({\boldsymbol x})$ is a diagonal matrix with main diagonal of ${\boldsymbol x}$;
 ${\bf I}_N$ is the $N\times N$ identity matrix; ${\bf 0}$ represents all-zero matrix with appropriate dimension.

\section{System Model}

As illustrated in Fig. 1, we consider a transmission system with a legitimate transmitter Alice equipped with $M$ antennas, a legitimate single-antenna user Bob, and an active eavesdropper Eve. We consider a TDD communication system, where the downlink channels and uplink channels are assumed to be reciprocal. The uplink reverse training phase is
 a typical way for Alice to obtain the channel information from Bob in order to apply beamforming in downlink data transmission.
We consider the training pilot signals are repeatedly used and publicly known, which allows the smart eavesdropper Eve be able to transmit the same training signals to confound Alice.

The channels from Bob and Eve to Alice are respectively modeled as the following length-$M$ vectors:
$ {\bf h} = [ h(1), h(2), \cdots, h(M) ]^T$ and
${\bf g} = [ g(1), g(2), \cdots, g(M) ]^T$.
We assume each element of ${\bf h}$ and ${\bf g}$ follows i.i.d. complex Gaussian distribution with variance $1/M$, such that the total average channel gains from Bob and Eve at all receive antennas are normalized, i.e., $E[\|{\bf h}\|^2]=1$ and $E[\|{\bf g}\|^2]=1$.

Denote ${\bf s}=[s(1),s(2),\cdots, s(N)]^T$ as the publicly known training sequence transmitted from  Bob. Assume the symbols $s(n)$ are drawn from constant modulus constellations, i.e., $|s(n)|=1$.
Denote the transmit power of Bob and Eve by $P_B$ and $P_E$, respectively.
We define two events, $\mathcal{H}_0$ and $\mathcal{H}_1$; Namely, $\mathcal{H}_0$: there exists no active eavesdropper who conducts pilot contamination attack;  $\mathcal{H}_1$: the active eavesdropper conducts pilot contamination attack trying to steal the information from the transmitter.
Then, under perfect frequency synchronization, the received signal at Alice during the uplink training period can be expressed as the following $N\times M$ matrix:
\begin{align}
\mathcal{H}_0: & \quad {\bf Y} =  \sqrt{P_B} {\bf s} {\bf h}^T  + {\bf N}, \label{equ1} \\
\mathcal{H}_1: & \quad  {\bf Y} = \sqrt{P_B} {\bf s} {\bf h}^T +  \sqrt{P_E} {\bf s} {\bf g}^T  + {\bf N} = \sqrt{P_B}{\bf s} {\bf h}_{\rm eq}^T + {\bf N}, \label{equ2}
\end{align}
where ${\bf h}_{\rm eq} = {\bf h}^T + \sqrt{\frac{P_E}{P_B}}{\bf g}^T $ and ${\bf N}\in\mathbb{C}^{N\times M}$ denotes the corresponding additive white Gaussian noise (AWGN) matrix. Assume each element of ${\bf N}$ follows i.i.d. complex Gaussian distribution with variance $\sigma_n^2$.
It is clear that in the case of Eve's attack, Alice obtains a composite equivalent channel ${\bf h}_{\rm eq} $ instead of the expected ${\bf h}$.
If Alice is unaware of attack and employs ${\bf h}_{\rm eq} $ as the legitimate channel information in the downlink beamforming design, this not only degrades the signal reception quality at Bob but also leads to a significant signal leakage to Eve
during the subsequent downlink data transmission.

\section{Proposed Training Phase with UFS}

From the comparison between (\ref{equ1}) and (\ref{equ2}), since Eve acts exactly in the same way as Bob,
it is in fact quite difficult for Alice to distinguish whether Eve is presence or not.
Motivated by the fact that frequency asynchronous nature would result in divergence between the transmitted signals from Bob and Eve, and thus may facilitate the detection of pilot contamination attack, we propose a new UFS scheme in this section to detect the pilot contamination attack.

Without loss of generality, we assume the length-$N$ training sequence block can be divided into $K$ equal-length segments, each with $Q$ symbols, i.e., $N=K\times Q$.
We can then split the whole training sequence block ${\bf s}$ into
\begin{align}
{\bf s} = \Big[ {\bf s}_1^T, {\bf s}_2^T,\cdots, {\bf s}_K^T\Big]^T,
\end{align}
where ${\bf s}_k = \big[ s(1\!+\!(k\!-\!1)Q), s(2\!+\!(k\!-\!1)Q), \cdots, s(kQ) \big]^T \in\mathbb{C}^{Q\times 1}$ denotes the pilot symbols in the $k\rm{th}$ segment.
In the UFS scheme,  Bob sends the pilot signals deliberately with some random CFO-like phase shifts. Specifically, let Bob independently select $K$ artificial CFOs for each training segment, namely $\Delta f_{B,k}$, $k=1,2,\cdots,K$.
  The normalized CFO of Bob in the $k\rm{th}$ segment can be expressed as $\phi_{B,k} = \Delta f_{B,k}T_s$, where $T_s$ denotes the symbol rate interval.
  Thus, $2\pi \phi_{B,k}$ in fact represents the deliberately introduced phase shift of Bob over the consecutive pilot symbols in the $k\rm{th}$ training segment.
  On the other side, since Eve has no knowledge about the random artificial CFOs of Bob, she can also randomly select one trial CFO for each segment, denoted by $\Delta f_{E,k}$, $k=1,2,\cdots,K$.
 Correspondingly, the normalized CFO of Eve in the $k\rm{th}$ training segment can be expressed as $\phi_{E,k} = \Delta f_{E,k} T_s$.

Let
${\bf E}(\phi) = \textrm{diag}(1, {\rm e}^{{\bf j}2\pi \phi}, \cdots, {\rm e}^{{\bf j}2\pi \phi (N-1) }   )$
 stand for the diagonal matrix representing the phase shift introduced by the CFO $\phi$.  The equivalent transmitted pilot symbols from Bob and Eve in the $k\rm{th}$ segment can be expressed as ${\bf E}(\phi_{B,k}){\bf s}_k$ and ${\bf E}(\phi_{E,k}){\bf s}_k$, respectively.
Then, the received signals at Alice in the $k\rm{th}$ segment can be expressed as the following $Q\times M$ matrix:
\begin{align*}
& \mathcal{H}_0: \quad {\bf Y}_k \!=\! \sqrt{P_B} {\bf E}(\phi_{B,k}){\bf s}_k {\bf h}^T + {\bf N}_k, \\
& \mathcal{H}_1: \quad {\bf Y}_k \!=\! \sqrt{P_B} {\bf E}(\phi_{B,k}){\bf s}_k {\bf h}^T\!\! +\!\!  \sqrt{P_E} {\bf E}(\phi_{E,k}){\bf s}_k {\bf g}^T   \!\!+\! {\bf N}_k,
\end{align*}
where ${\bf N}_k\in\mathbb{C}^{Q\times M}$ denotes the corresponding AWGN matrix, and each element of ${\bf N}_k$ is i.i.d. complex Gaussian variable with variance $\sigma_n^2$.

The attack detection algorithms at Alice based on the received signals ${\bf Y}_k$, $k=1,2,\cdots,K$, will be developed in the following sections. Next, we first consider the CFO and channel estimation issue without Eve's attack. Specifically, when Eve is absent, the maximum-likelihood (ML) single-user CFO estimation and channel estimation at Alice can be performed  as follows:
\begin{align}\label{equ3}
\{\hat\phi_{B,k}, \hat{\bf h} \} =  \arg\min_{\tilde\phi_{B,k}, \tilde{\bf h}} \sum_{k=1}^K \Big \| {\bf Y}_k - {\bf E}(\tilde\phi_{B,k} ) {\bf s}_k \tilde{\bf h}^T   \Big\|^2,
\end{align}
where $\tilde\phi_{B,k}$ and $\tilde{\bf h}$ stand for the trial CFO and channel for Bob, respectively. The CFO estimation solution of (\ref{equ3}) can be expressed as
\begin{align}\label{equ4}
\hat\phi_{B,k} =  \arg\max_{\tilde\phi} \big \| {\bf s}^H{\bf E}(\tilde\phi)^H {\bf Y}_k  \big\|^2,
\end{align}
which targets at maximizing the correlation between the pilot and the received signal after CFO compensation of a trial CFO $\tilde\phi$.
 The channel estimation solution can be then given by
 \begin{align}\label{equ10}
 \hat{\bf h}^T = \frac{1}{N}  \sum_{k=1}^K {\bf s}^H{\bf E}(\hat\phi_{B,k} )^H {\bf Y}_k.
 \end{align}

Regarding the CFO and legitimate channel estimation performance of UFS scheme, we obtain the following \textit{Lemma}:

\textit{Lemma 1:} In the proposed UFS scheme, the CFO and channel estimation mean square error (MSE) under the high signal-to-noise ratio (SNR) condition in the absence of Eve's attack can be expressed as:
\begin{align}
& \textrm{MSE}(\hat\phi_{B,k}) =  \frac{3\sigma_n^2}{2\pi^2 Q(Q^2-1)\|{\bf h}\|^2 }, \\
& \textrm{MSE}(\hat{\bf h}) = \frac{\sigma_n^2}{N}\left( \frac{3}{8\pi^2 } \frac{Q-1}{Q+1} + M \right). \label{lemma1equ1}
\end{align}
The detailed proof is omitted due to the space limitation.
The following observations can be made from \textit{Lemma 1}:
First, as expected, the CFO estimation performance highly depends on the length of each training segment. This says that for better CFO estimation performance, a single training segment is preferred, that is, $K=1$ and $Q=N$. However, this may not be a good choice for purpose of attack detection, which will be discussed in the following sections. Nevertheless, we see that the channel estimation performance, which is the main concerned issue in our work, is basically irrelevant to the segment length. In other words, the legitimate channel estimation performance is insensitive to the number of segments.

Second, the corresponding MSE performance without artificial frequency shifts in the conventional frequency synchronous channel estimation  can be obtained as $M\sigma_n^2/N$.
As compared to this conventional benchmark, the relative increased MSE in channel estimation of our UFS scheme can be expressed as
$ \frac{1}{M} \frac{3}{8\pi^2 } \frac{Q-1}{Q+1} <\frac{1}{25M}$.
 The above discussions indicate that the multiple artificial frequency shifts in our UFS scheme introduce only negligible performance degradation in legitimate channel estimation, especially with more receive antennas at Alice. This differs from the existing superimposed random sequence based scheme~\cite{TugnaitWCL} which suffers from a lot performance degradation in terms of the legitimate channel estimation.

\section{Proposed UFS-MDL Detection Algorithm }

In this section, we develop a source enumeration based algorithm, referred to as `UFS-MDL', to detect the presence of attack at Alice based on the received signals ${\bf Y}_k$, $k=1,2,\cdots,K$. The corresponding asymptotic detection performance analysis is also provided.

\subsection{Proposed Detection Scheme}

Owing to the uncoordinated frequency shifts, our UFS training scheme introduces the
divergence between the equivalent transmitted pilot signals between Bob and Eve. Hence,
for the case $M\ge 3$, the minimum description length (MDL) algorithm~\cite{Wax85} can be employed by Alice to detection the presence of Eve.

Specifically, the autocorrelation matrix for the $k\rm{th}$ training segment can be expressed as
\begin{align}
{\bf R}_k = \frac{1}{Q}E\Big[ {\bf Y}^T_k {\bf Y}^*_k \Big].
\end{align}
In the absence of attack, there holds
\begin{align}
\mathcal{H}_0: \hspace{2mm} {\bf R}_k = P_B {\bf h}{\bf h}^H + \sigma_n^2 {\bf I}_M,
\end{align}
which implies the dimension of signal subspace of ${\bf R}_k$ is only one.

Let $\Delta\phi_k = \phi_{B,k}-\phi_{E,k}$ represent the CFO between Bob and Eve in the $k\rm{th}$ training segment.
In the presence of Eve's attack, we have
\begin{align}
\mathcal{H}_1: \hspace{2mm}  {\bf R}_k = & [ {\bf h}, {\bf g} ]
\underbrace{ \left[
\begin{array}{cc}
P_B & \sqrt{P_BP_E} \rho_k \\
\sqrt{P_BP_E} \rho_k^*  & P_E
\end{array}
\right] }_{ {\bf P}_k }
[ {\bf h}, {\bf g} ]^H \nonumber \\ & + \sigma_n^2 {\bf I}_M.
\end{align}
 Here, ${\bf P}_k\in\mathbb{C}^{2\times 2}$ represents the correlation of the equivalent transmitted pilot signals between Bob and Eve, where
\begin{align}\label{equ7}
\rho_k = \frac{1 - {\rm e}^{{\bf j}2\pi Q\Delta\phi_k } }{Q( 1-{\rm e}^{{\bf j}2\pi\Delta\phi_k } ) }.
\end{align}

Theoretically, when $\Delta\phi_k\ne 0$, we have
\begin{align}
\det\left( {\bf P}_k \right)  = P_BP_E (1 - |\rho_k|^2) >0,
\end{align}
saying ${\bf P}_k$ is nonsingular in this case. The dimension of signal subspace of ${\bf R}_k$ is two when $\Delta\phi_k\ne 0$.

Following the above discussions under $\mathcal{H}_0$ and $\mathcal{H}_1$,
the MDL criterion~\cite{Wax85} can be employed to determine the dimensional of signal subspace. Specifically, denote the eigenvalues of $\hat{\bf R}_k = \frac{1}{Q}{\bf Y}_k^T{\bf Y}^*_k$ in descending order by $\lambda_k(i)$, $i=1,2,\cdots,M$. The MDL estimation of signal subspace dimension for $\hat{\bf R}_k$ can be given by
\begin{align}
\hat{d}_k = \arg\min_{1\le d\le M-1} \mathbb{MDL}_k(d),
\end{align}
where
\begin{align}
\mathbb{MDL}_k(d) \!\!=\!\! & -\!\!\sum_{i=d+1}^M  \log( \lambda_k(i)) \! + \! (M\!-\!d)\log\Big( \frac{\sum\limits_{i=d+1}^M \!\!\lambda_k(i)  }{M-d} \Big)  \nonumber \\ & \kern 50pt + \frac{d(2M-d)\log(Q)}{2Q}.
\end{align}

When the signal subspace dimension of at least one training segment is above one, i.e., $\max\limits_k \hat{d}_k >1$, we declare the presence of Eve's attack. Otherwise, we consider the Eve's attack is absence.

\subsection{Asymptotic Performance Analysis}

In the subsection, we provide asymptotic performance analysis for the proposed UFS-MDL detection algorithm. We consider the high SNR condition. Moreover, both the training sequence length and the receive antenna number at Alice are assumed to be very large, i.e, $N\gg 1$ and $M\gg 1$. Note that there is an increasing interest from both academy and industry to equip base station with a large scale antenna array~\cite{GaoTVT16,GaoAccess16}, such a system can provide a remarkable increase in both reliability and spectral efficiency.
We focus on the asymptotic false negative (miss detection) probability, which is defined as the probability when miss of detection happens.
The following \textit{Lemma} can be obtained:

\textit{Lemma 2:} Assume the random normalized CFOs of Bob and Eve follow uniform distribution from $-\phi_{\max}$ to $\phi_{\max}$ with $|\phi_{\max}|<0.5$. Denote $P_{\rm th} = M(\sqrt[Q]{Q}-1) \sigma_n^2$. When the minimum power of Bob and Eve is larger than $P_{\rm th}$, i.e., $\min(P_B,P_E) > P_{\rm th}$, the asymptotic miss detection probability of the UFS-MDL detection algorithm can be approximately upper bounded by
\begin{align}\label{equlemma3}
 \mathcal{P}_{{\rm MDL} } =  \left(  \frac{2}{\pi Q \phi_{\max}}
 \arccos \sqrt{ \left(1-\frac{P_{\rm th}}{P_B} \right) \left(1-\frac{P_{\rm th}}{P_E} \right)  } \right)^K.
 \end{align}
 Otherwise, when $\min(P_B,P_E) \le P_{\rm th}$, the miss detection probability of UFS-MDL  approximately equal one, i.e., $\mathcal{P}_{{\rm MDL} }\lesssim 1$.
The detailed proof is omitted due to the space limitation.
 The following observations can be made from \textit{Lemma 2}:
1) As expected, the miss detection probability of UFS-MDL can be reduced by increasing the range of random artificial CFOs. The asymptotic miss detection probability is also relevant to the noise power and the number of receive antennas. There exists a cliff-like jump of miss detection probability when the minimum power between Bob and Eve becomes less than $P_{\rm th}$.
2)  We see that the detection performance of UFS-MDL can always benefit from the increased attack power from Eve.
 Interestingly, the detection performance will asymptotically touch a lower bound when attack power $P_E$ becomes very large. Specifically, when $P_E\gg P_B$,
according to (\ref{equlemma3}), the asymptotic lower bound of UFS-MDL can be obtained as
\begin{align}\label{equ8}
 \mathcal{P}_{{\rm MDL} } \gtrsim \left(  \frac{2}{\pi Q \phi_{\max}}
 \arccos \sqrt{ \left(1-\frac{P_{\rm th}}{P_B} \right)  } \right)^K.
\end{align}
It is observed that this lower bound can be simply reduced by a larger transmit power at Bob.
3) The equation (\ref{equlemma3}) also demonstrates the benefit from the multiple frequency shifts in UFS-MDL.

\section{Simulations}

In this section, we provide numerical results to evaluate the performance of the proposed scheme. Assume the pilot symbols are randomly drawn from QPSK constellation.
Unless otherwise specified, the pilot length is taken as $N=64$  and let normalized artificial CFOs of Bob and Eve in each training segment follow uniform distribution from -0.2 to 0.2, i.e., $\phi_{\max}=0.2$.

\begin{figure}[t]
\begin{center}
\includegraphics[width=85mm]{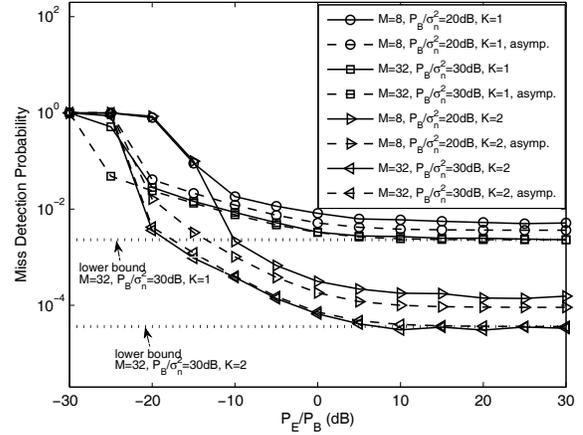}
\end{center}
\vspace{-5mm}
\caption{ Miss detection probability of the proposed UFS-MDL algorithm with different relative power between Bob and Eve.  The corresponding asymptotic analytical performance is included as the dashed curves, while the asymptotic lower bound in (\ref{equ8}) is plotted as the dotted curves. }
\end{figure}

\begin{figure}[t]
\begin{center}
\includegraphics[width=85mm]{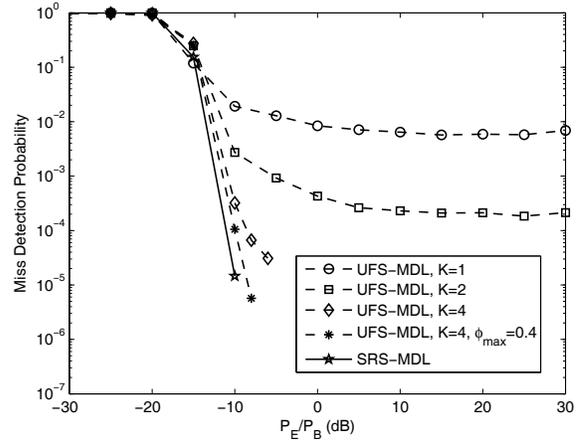}
\end{center}
\vspace{-5mm}
\caption{ Detection performance comparison between the SRS-MDL scheme~\cite{TugnaitWCL} and the proposed UFS scheme ($M=16$, $P_B/\sigma_n^2=$20 dB). }
\end{figure}


In Fig. 2, we show miss detection probability of our UFS-MDL algorithm as a function of the relative power between Bob and Eve ($P_B/P_E$) under different SNR condition $P_B/\sigma_n^2$.
The following observations can be made.
First, as expected, the performance of UFS-MDL can be improved with stronger signal power from Eve.
Second, we see that the simulation results closely match the corresponding asymptotic results, which verifies the correctness of the asymptotic analysis. Especially, when Eve has much stronger power than Bob, we can observe that the detection performance converges to the analytical lower bound given by (\ref{equ8}).

The detection performance of UFS-MDL with different $K $ is plotted in Fig. 3. We also include the performance of the superimposed random sequence based MDL scheme~\cite{TugnaitWCL}, labelled as `SRS-MDL'.
 The results show that when Eve has smaller or similar power as Bob,
  the proposed scheme can achieve comparable detection performance as SRS-MDL with a relative large $K$ and a larger maximum artificial CFO $\phi_{\max}$.
 However, SRS-MDL behaves  better in the region with much stronger attack power.
   Note that the performance of UFS-MDL approaches a lower bound as the attack power increases.
Nevertheless, as the random sequences are superimposed in the pilot signal in SRS-MDL, our UFS scheme could substantially outperform SRS-MDL in terms of channel estimation, which will be demonstrated below.

\begin{figure}[t]
\begin{center}
\includegraphics[width=85mm]{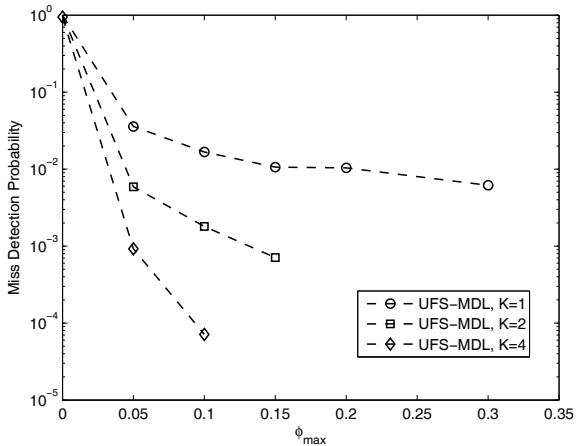}
\end{center}
\caption{ Detection performance of the proposed UFS scheme with different maximum artificial CFOs ($M=16$, $P_B/\sigma_n^2=20$dB and $P_B=P_E$). }
\end{figure}

\begin{figure}[t]
\begin{center}
\includegraphics[width=85mm]{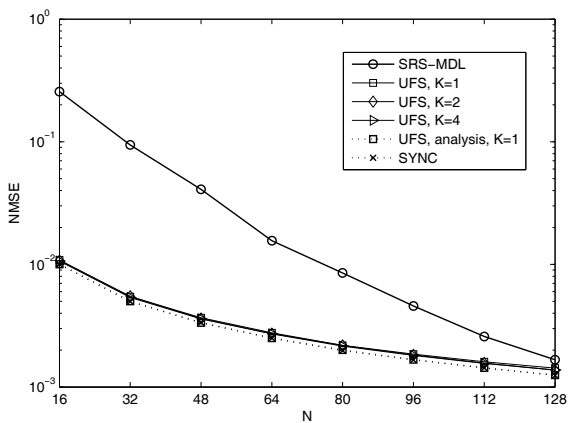}
\end{center}
\caption{ Channel estimation performance comparison between the SRS-MDL scheme~\cite{TugnaitWCL} and the proposed UFS scheme. }
\end{figure}


In Fig. 4, we increase the maximum artificial CFO $\phi_{\max}$ from 0 to 0.3 and demonstrate the  detection performance evolution of our UFS scheme. As expected, we see that the detection performance of our proposed UFS scheme can be improved by increasing the maximum artificial CFOs.
The results also clearly indicate the benefit of multiple segments.
It is seen that, in the case of more segments, i.e., a larger $K$, more random artificial CFOs are introduced and the miss detection probability could decline more quickly as the maximum CFO increases.

In the last, we plot the channel estimation performance comparison between SRS-MDL  and our UFS scheme in Fig. 5. The MSE of channel estimation is adopted as the figure of merit. We assume absence of Eve in this example. The iterative channel estimation method is employed in SRS-MDL as described in~\cite{TugnaitWCL}.
It is seen that the channel estimation performance of both SRS-MDL and our scheme improves with increasing the pilot signal length. Moreover, our scheme show no much changes with different $K$.
Note that SRS-MDL superimposes the self-contamination random sequences in the pilot signal, inevitably degrading the performance of legitimate channel estimation.
In comparison, our scheme could outperform SRS-MDL especially with a shorter pilot sequence.
On the other side, we include the corresponding analytical results from (\ref{lemma1equ1}) with $K=1$ in this figure, plotted as the dotted curve.  The conventional channel estimation performance with frequency synchronization is also included, labelled as `SYNC'. It is observed that the simulation results of our scheme closely approach the analytical curve and the SYNC benchmark. This coincides with our previous observation that the proposed UFS scheme basically does not sacrifice the legitimate channel estimation performance for attack detection.

\section{Conclusions}

In this paper, we proposed a new UFS scheme for detection of pilot contamination attack. The proposed scheme deliberately introduces multiple random frequency shifts in the transmitted pilot signal from the legitimate user Bob. A detection algorithms were designed for Alice to detect the presence of attack. We also provided both the asymptotic performance analysis and numerical results to verify the proposed studies.

\end{document}